# Benchmarking the ONOS Intent interfaces to ease 5G service management

Rami Akrem Addad[1], Diego Leonel Cadette Dutra[1], Miloud Bagaa[1], Tarik Taleb[1]
Hannu Flinck[2], and Mehdi Namane[1]
[1] Dep. of Communications and Networking School of Electrical Engineering, Aalto University, Espoo, Finland
[2] Nokia Bell Labs, Espoo, Finland
Emails:{firstname.lastname}@aalto.fi; hannu.flinck@nokia-bell-labs.com

*Abstract*—The use cases of the upcoming $5G$ mobile networks introduce new and complex user demands that will require support for fast reconfiguration of network resources. Software Defined Network (SDN) is a key technology that can address these requirements, as it decouples the control plane from the data plane of the network devices and logically centralizes the control plane in the SDN controller. SDN network operating system (ONOS) is a state-of-art SDN controller that aims to address this important scalability limitation from its design. An important feature of ONOS is that it allows network administrators to configure and manage networks with a high-level of abstraction by using Intent specifications. An Intent is a policy expression describing what is the desired outcome rather than how the outcome should be reached. The concept of Intents coupled with the distributed storage space are the key components for the theoretical scalability of ONOS. In this paper, we present our evaluation of the ONOS Intent northbound interface using a methodology that takes into consideration the interface access method, type of Intent and number of installed Intents. Our preliminary analysis indicates a linear increase in the computational cost with regards to the number of submitted Intents, with the access method being a major factor in the overall computational cost.

## I. INTRODUCTION

5G networks are not only about broadband connectivity but are also intended to create a basis for the provisioning of a wide variety of applications that have performance demands beyond the capacity of current networks. In order to meet the user expectations and anticipated traffic growth, mobile network operators are overhauling their communication infrastructure to $5G$ [1]. Equally important to investments in new radio and transport technologies are the investments to manage and control the network resources. The Software Defined Networking (SDN) [2], [3] technology brings flexibility in the granularity of controlling network resources [4], [5]. It is commonly identified as a key enabler for $5G$ networks since it can efficiently provide on-demand reconfiguration of the critical networking resources [6], [7]. SDN control may be based on OpenFlow [8], NETCONF, PCEP, and similar protocols that separate traffic flow processing from the control and management of the network devices performing the traffic flow processing (e.g. routing or switching).

To take advantage of the flexibility of the SDN concept, a new network element, SDN controller [9]–[12], needs to be deployed. The controller is a logically centralized entity that is responsible for storing the current network configuration requested by its operator and to program this configuration to the SDN-enabled switches of the infrastructure. The controller may also be involved in traffic flow processing as the switches may request it to handle packets that they do not know how to forward. The OpenFlow protocol supports such reactive operation mode in SDN-enabled networks, which facilitates automatic reconfiguration of the network. This fine-grained level of control is typically implemented by embedding or linking an application with the controller software or using a RESTful interface offered by the controller.

While the first generation of SDN controllers were both logically and physically centralized software, this introduced major concerns about the scalability of the SDN-enabled networks. This was addressed by a new generation of controllers, e.g., OpenDaylight, that aimed to distribute the controller over multiple servers. However, this approach led to the known problems of consistency of distributed systems [13], [14]. The OpenDaylight [15] project chose to address this issue of a classical trade-off between scalability and implementation by using a strong consistency model that reduces the implementation burden but also limits scalability, while still providing high availability.

The next generation of SDN controllers aims to implement a higher level abstraction to manage the network functionality through their northbound interfaces by hiding the details of the OpenFlow and other protocols from the operator. The Open Network Operating System (ONOS) [16] is a state-of-art SDN controller that was designed to guarantee scalability, high-availability, and performance. It offers the possibility to configure and manage networks in a simple way, leveraging a high-level abstraction provided by ONOS Intents. Conceptually, Intents represent *desires* of the network administrator what to do rather than how to do. ONOS compiles Intents into one or more Flow Rules that implement the desired network behavior.

While previous works have benchmarked the performance of other SDN controllers, most of them used simulation or emulation tools, such as Cbench. These tools aimed to generate an OpenFlow traffic, which is equivalent to benchmarking ONOS' southbound interface [17]. Since the Intent interface of ONOS represents a major feature in its scalability, it is important to understand in detail how the use of Intents impacts the performance of the controller as a function of the Intent load submitted to it. It is equally important to understand if there are any performance differences between the different methods of providing the Intent specifications. The major contributions of this paper are an experimental evaluation of the performance of ONOS RESTful and command line northbound interfaces for the 3 basic Intent types, and a preliminary analysis based on these results that point to best practices in the development of complex behaviors and ONOS-aware distributed applications.

The remainder of this paper is organized as follows. Section II presents related work. In Section III, we describe the benchmarking methodology that we used for evaluating ONOS. In Section IV, we present the results of our experimental evaluation. Finally, Section V concludes the paper and introduces some future research work.

## II. BACKGROUND AND RELATED WORK

### A. Related Work

In recent years, SDN has gained a lot of momentum with a number of open sources controllers being developed for various purposes. In this section, we focus on the research work focusing on ONOS and its performance.

Berde et al. [16] analyze the performance of the first two ONOS prototypes, introducing the path installation performance metric for the second prototype. The metric covers the average time of processing application requests up to the network state update, showing how quickly the physical network is configured. Kim et al. [18] presented the design and the implementation of OFMon that is a monitoring system recording OpenFlow message exchanges on the ONOS logging system. It provides in real time monitoring results to the network administrator through both the ONOS command line interface (CLI) and the GUI. We use this tool for the performance analysis of CPU and memory consumption.

Yamei et al. [17] presented a comparative analysis of the performances of both ONOS and OpenDaylight [15] Controllers using three tools: IXIA Test Instrument(XG12), Cbench, and Mininet. Those tests were done measuring metrics like channel capacity, flow modification delay and cluster fail-over time. The Open Networking Laboratory [19] has done a number of detailed performance analysis and scale-out tests such as topology event latency, Intents processing latency, and topology scaling operation.

Salman et al. [20] presented a deep performance comparison of a number of SDN controllers including ONOS, OpenDaylight, and LIBFLUID_RAW. The comparison was done using Cbench, the most commonly used OpenFlow testing tool. The SDN controllers were evaluated using both the latency and the throughput modes from Cbench, varying other parameters such as the number of switches and threads. Bianco et al. [21] developed a quantitative model in order to estimate the number of exchanged packets on the control plane when setting up flows between ONOS and the switches it controls. They applied their model to different realistic network topologies. They also evaluated precisely the OpenFlow traffic generated on the control plane and the average number of flow rules installed in each switch.

### B. Background

*1) OpenFlow:* OpenFlow [8] was the first protocol that implemented the SDN concept of extracting switch control to a centralized controller. OpenFlow provides a detailed view and control of the traffic flows passing through SDN-enabled switches. In this way, OpenFlow bypasses the restrictions of a switch control that is "locked" into proprietary operating systems of network equipment. As a result, network operators would be able to implement new strategies and protocols for traffic forwarding and routing.

An OpenFlow-enabled switch [22] includes a flow table that contains rules for identifying different types of traffic, processing each packet of the flow (forwarding, dropping, adding/modifying VLAN tags), and recording the number of packets processed for this flow. If no rule matches to a packet, the switch can send the header of the packet to the SDN controller. The latter would make further decisions on how the packet should be processed and pushes appropriate entries to create flow rules into the flow table of the switch. These new rules will be used for the similar subsequent packets. Thanks to this strategy, the packet processing overhead on the SDN controller should be mitigated.

*2) The Open Network Operating System:*

**O**pen **N**etwork **O**perating **S**ystem – ONOS – is an SDN controller that has been conceived to support scalability, high availability and performance [23]. It was designed for network service providers with the following key features: a distributed core for maximum scalability, modularity, southbound abstractions, and northbound abstractions. The latter is divided into two main parts, the global network view (which provides the applications with a view of the network - the hosts, switches, and links), and the Intent Framework, that enables the network administrator to manage the network with a high-level of abstraction by submitting Intents. An Intent could be, for example, to set up a connection between two particular hosts in the network. Intents will be handed over to the ONOS core, which is responsible for translating them via *Intent Compilation*, into *Installable Intents*, which are *Actionable Operations* to ONOS. The compilation process takes into account the network state and behavior requested in the Intent to generate one or more flow rules to implement the requested behavior in the ONOS managed network. These actions are then carried out by the *Intent Installation* process, which results in a set of flow rules being installed on one or more selected switches in the network.

As shown in Figure 1, there are three different ways to push Intents on ONOS: **(1)** by creating an ONOS application, **(2)** using the Command Line Interface (CLI), or **(3)** through the RESTful interface. The first two access methods are equivalent, which means that the CLI will delegate the Intent creation according to the ONOS application; while using the RESTful interface is equivalent to calling a Web Service with a particular request using the HTTP protocol. This Web Service will relay this request to the ONOS Core, after having transformed it into a request in a query format that can be understood by the latter. The generic basic Intents provided by ONOS are: *point-to-point*, *single-to-multi-point*, *multi-to-single-point* and *host-to-host*. We have chosen to benchmark the performances of the three first Intents, as for our test network the *host-to-host* Intent would not provide useful insights to ONOS Intent framework as it will be translated into a set of *point-to-point* Intents.

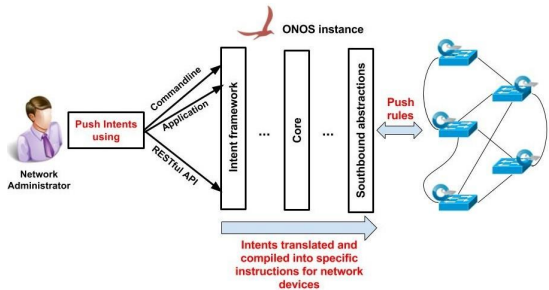

Fig. 1. ONOS Intent framework mechanism.

### III. METHODOLOGY

Evaluating SDN controllers based on simulated or emulated traffic sets is effective to better understand their strength, robustness, and scalability. However, such an approach is not suitable for benchmarking the Intent abstraction of ONOS. While the benchmarking of the southbound interface can be done using Cbench (as discussed in Section II) to generate *flow requests*, the northbound interface that implements Intent abstraction operates not on flows but policy-based directives containing descriptions of network resources, their constraints and criteria to select traffic.

To the best of our knowledge, the other published performance evaluation of the SDN controllers used simulation tools to create flow requests with the focus mainly on the interaction between the SDN controller and the Open Flow protocol. Instead, our work is focusing on how the SDN controllers behave and how their performance changes when using different Intents and what is the impact of using different interfaces to communicate Intents to the ONOS controller. Moreover, we focus our evaluation on the basic Intents provided by the ONOS core architecture, as we study how ONOS behaves when an external application is submitting Intent installation requests at a high rate. The results presented in this paper allow us to evaluate how an external application can use ONOS to install its desired network configuration while mitigating the cost associated with the communication with the SDN controller.

In our benchmarking of the Intent interface, we are interested in how ONOS behaves in terms of agility and scalability. We define **Agility** as the amount of time required to successfully install a set of Intents, and **Scalability** as how many Intents a single ONOS instance can manage. These two metrics provide a clear understanding of the expected performance of an ONOS deployment to a network administrator. For our analysis, we have selected these three basic Intent types *point-to-point*, *single-to-multi-point*, and *multi-to-single-point* that would be used to create more complex Intents.

The compilation and installation times of an Intent depends significantly on the used computational resources, e.g., CPU cycles and memory of the used servers. Therefore, to make our results independent from deployed testbed configuration, we selected to measure how long it takes to install a fixed set of Intents using both the CLI and the RESTful interfaces provided by ONOS. For the CLI, we augmented the ONOS code, introducing timekeeping mechanism inside the ONOS core for each CLI Java class, in our case: **(1)** *AddPointToPointIntentCommand.java*, **(2)** *AddSinglePointToMultiPointIntent.java*, and, **(3)** *AddMultiPointToSinglePointIntentCommand.java* that are used to create *point-to-point*, *single-to-multi-point*, and *multi-to-single-point* Intents respectively. We also measured the time for the add method of each Intent we were benchmarking. For the RESTful interface tests, we measured the time spent to push the benchmarked Intents through python scripts.

### IV. EXPERIMENTAL EVALUATION

In this section, the results of our evaluation of the basic Intent types over Intent interfaces of ONOS are shown. The testbed used in the evaluation consists of two computer nodes with configurations as in Table I. We are using version **1.9.0** of **ONOS**. Our test network topology is built using Open vSwitchs (OVS).

The evaluation of the CLI was carried out directly on the server running our ONOS instance, while we use another computer to submit the RESTful requests through the network to ONOS. The configuration of this auxiliary computer is described in Table II.

TABLE I
SERVER CONFIGURATION.

| Component | Description |
|---|---|
| CPU | Intel(R) Xeon CPU E5-2640 v3 *2.60GHz* |
| RAM (GB) | 8GB |
| Network | 1Gbps |
| OS (Version) | Ubuntu 16.04 LTS |

For each type of benchmarked Intent type– *point-to-point*, *multi-to-single-point*, and *single-to-multi-point*, we ran 50 iterations to find out the mean installation times, standard deviations, the 95% confidence interval for each set of the Intents. We vary the number of Intents (workload) installed in each experiment from 1,000 to 20,000. This procedure was adopted to mitigate possible noise that could be introduced by the processor cache, system daemons, Java interpreter, or the network jitter. For the results plotted henceforth, the Y-axis shows the total time to install the Intents in milliseconds, while on the X-axis, we show the number of installed intents in the experiment. The figures show also the mean and the 95% confidence intervals.

TABLE II
AUXILIARY COMPUTER NODE: LOAD GENERATOR FOR THE RESTFUL INTERFACE.

| Component | Description |
|---|---|
| CPU | Intel(R) core(TM) i5-6300U @ 2.40GHz |
| RAM (GB) | 8 GB |
| Network | 1Gbps |
| OS (Version) | Ubuntu Desktop 16.04 LTS |

### A. Multi-to-Single-Point Intent

Figure 2 summarizes our results for applying *Multi-to-Single-Point* Intents through CLI and RESTful interfaces of ONOS varying workload from 1,000 to 20,000 Intents. The 95% confidence interval and the mean are calculated using 50 executions of the same Intent. As the 95% confidence intervals were very small with respect to the mean values for both the CLI and the RESTful interface, we magnified the confidence intervals 50 times in the figure (instead of showing the original values). As can be seen from the figure, the installation time through ONOS RESTful interface grows from 637.990 $ms$ to 12,330.511 $ms$, when we increase the number of Intents from 1,000 to 20,000 and there is a 20 fold increase in the mean installation time. Moreover, as we increased the number of installed Intents, our sample distribution became more stable reducing the coefficient of variation from 0.0127 to 0.008, which suggests that we did not overload ONOS in this experimental scenario.

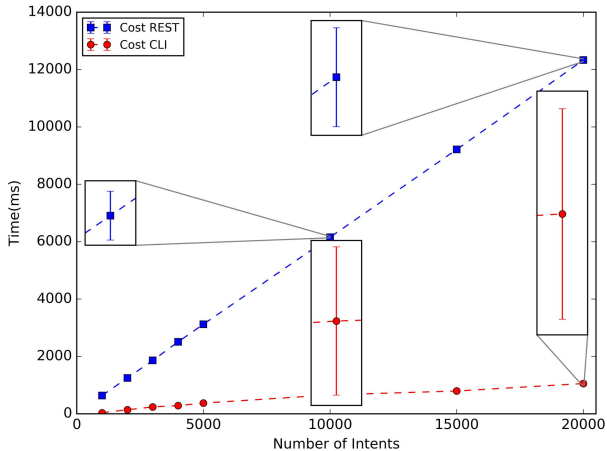

Fig. 2. Multi-to-Single-Point intent - benchmarking results.

For the CLI, Figure 2 shows also a linear increase in the installation time as we increase the number of Intents, albeit with a lower computational cost. The mean installation time for 1,000 intents was 35.771 $ms$ with a standard deviation and confidence of 13.686 $ms$ and 4.770 $ms$, respectively. For 20,000 Intents, the mean installation time is 1,054.343 $ms$ with a standard deviation of 210.661 $ms$. As the figure also shows, for 10,000 Intents, the mean installation time was 6,155.655 $ms$ and 658.857 $ms$ for the RESTful interface and CLI, while 95% confidence interval for them were 16.996 $ms$ and 51.652 $ms$, respectively.

TABLE III
MEAN INSTALLATION TIME AND ITS 95% C.I. FOR MULTI-TO-SINGLE INTENT.

| Num. Intents | Mean RESTful ($ms$) | 95% C.I RESTful | Mean CLI ($ms$) | 95% C.I CLI |
|---|---|---|---|---|
| 1,000 | 637.99 | 2.816 | 35.771 | 4.77 |
| 2,000 | 1,251.51 | 4.718 | 142.857 | 9.793 |
| 3,000 | 1,859.501 | 7.295 | 238.8 | 19.257 |
| 4,000 | 2,506.155 | 7.769 | 285.971 | 17.619 |
| 5,000 | 3,122.163 | 7.831 | 372.629 | 21.819 |
| 10,000 | 6,155.655 | 16.996 | 658.857 | 51.652 |
| 15,000 | 9,215.017 | 28.424 | 794.8 | 50.064 |
| 20,000 | 12,330.511 | 34.455 | 1,054.343 | 73.421 |

The difference between computational costs of RESTful and CLI interfaces seems to be stemming from the amount of I/O code that the Intent requests have to pass through. The differences in installation times shown in Table III also hint to that direction.

### B. Single-to-Multi-Point Intent

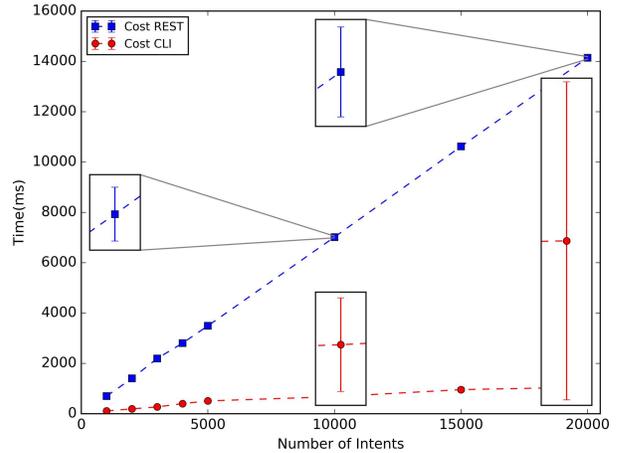

Fig. 3. Single-to-Multi-Point Intent - benchmarking results.

The evaluation of the *Single-to-Multi-Point* Intent is summarized in Figure 3 and in Table IV. As for the previous experiment, the mean values for both the RESTful and CLI interfaces are shown. There appears similar I/O overhead penalty for the RESTful interface as in earlier experimentation.

Also in Figure 3, we present the details of two workload points of our evaluation. For the RESTFul interface with $10,000$ Intents, the average execution time is $7,018.609\ ms$ with a confidence interval of $4.695\ ms$, whereas for the CLI the same workload took $678.171\ ms$ with a confidence interval of $37.218\ ms$. For the $20,000$ Intents workload, the mean installation times through the RESTFul interface and the CLI were $1,4143.2375\ ms$ and $1,055.7145\ ms$, respectively. The 95% Confidence Intervals for this experiment were $4.695\ ms$ and $35.7225\ ms$ for the RESTful interface and CLI, respectively.

TABLE IV
MEAN INSTALLATION TIME AND ITS 95% C.I. FOR SINGLE-TO-MULTI-POINT INTENT.

| Num. Intents | Mean RESTful (ms) | 95% C.I RESTful | Mean CLI (ms) | 95% C.I CLI |
|---|---|---|---|---|
| 1,000 | 700.589 | 4.695 | 109.800 | 13.900 |
| 2,000 | 1,403.436 | 5.817 | 192.514 | 17.544 |
| 3,000 | 2,196.356 | 4.081 | 270.600 | 31.681 |
| 4,000 | 2,805.094 | 5.822 | 401.657 | 40.376 |
| 5,000 | 3,493.791 | 7.800 | 509.057 | 49.977 |
| 10,000 | 7,018.610 | 21.372 | 678.171 | 37.218 |
| 15,000 | 10,619.652 | 24.334 | 957.886 | 86.654 |
| 20,000 | 14,143.237 | 35.722 | 1,055.714 | 126.397 |

*C. Point-to-Point Intent*

The two previous performance tests showed that the RESTful interface imposes a significant overhead on the installation of the intents. To evaluate further this behavior, we created a simulated complex intent that contains multiple Intents with just one RESTful call using the ONOS' *Point-to-Point* Intent, which implements a simple forwarding from one OVS port to another. Figure 4 presents the summary of our results with 50 times magnification of the $20,000$ workload point for the RESTful and CLI interfaces. As can be seen from Figure 4, the behavior of the simulated complex Intent has a similar computational cost as the CLI *Point-to-Point*, albeit with a higher overhead caused by the RESTful access code and the new API control code.

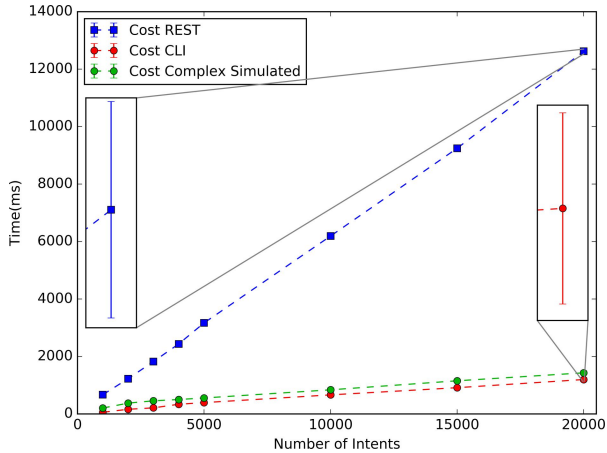

Fig. 4. "Point-to-Point" Intent - benchmarking results.

The mean installation time for $1,000$ Intents was $669.276\ ms$ for the RESTful interface and $57.2\ ms$ for the CLI, while the 95% confidence interval for the same experiments were $24.031\ ms$ and $9.468\ ms$, respectively. For $2,000$ and $3,000$ Intents, the mean installation time roughly increased at the same rate as the size of the workload for the RESTful interface as it increased from $1,226.505\ ms$ to $1,823.248\ ms$. The 95% confidence intervals were $11.712\ ms$ for $2,000$ and $20.107\ ms$ for $3,000$. The results for the CLI for the same workload follow the same trend, albeit with slower linear factor going from $162.228\ ms$ to $211.714\ ms$. This behavior is attributable to how our benchmark code interacts with ONOS. The evaluation of the CLI is based on an existing ONOS command that requests the ONOS core to execute an Intent installation multiple times. Therefore, our measurements take into account only the time spent by ONOS to install the Intent itself and not the time spent for the user I/O. However, for the RESTful interface, the installation of a workload was carried out through the ONOS default interface one by one per request.

TABLE V
MEAN INSTALLATION TIME AND ITS 95% C.I. FOR POINT-TO-POINT INTENT.

| Num. Intents | Mean RESTful (ms) | 95% C.I RESTful | Mean CLI (ms) | 95% C.I CLI |
|---|---|---|---|---|
| 1,000 | 669.276 | 24.032 | 57.2 | 9.468 |
| 2,000 | 1,226.505 | 11.712 | 162.229 | 18.225 |
| 3,000 | 1,823.249 | 20.108 | 211.714 | 20.682 |
| 4,000 | 2,436.431 | 17.329 | 337.343 | 37.725 |
| 5,000 | 3,167.603 | 40.439 | 394.4 | 42.135 |
| 10,000 | 6,192.263 | 77.685 | 663.829 | 64.341 |
| 15,000 | 9,246.514 | 104.522 | 913.886 | 45.687 |
| 20,000 | 12,622.108 | 75.340 | 1198.086 | 66.572 |

Table V shows the precise mean values plotted in Figure 4. The results offer further support to our interpretation of the different overhead increase rates of the RESTful and CLI interfaces. Finally, in Figure 4, we also present the results for the $20,000$ Intents workload for both interfaces. The mean installation time was $12,622.108\ ms$ with a confidence interval of $75.340\ ms$ for the RESTful interface, while for the CLI the mean installation time and confidence interval were $1,198.086\ ms$ and $66.572\ ms$, respectively.

*D. Result Discussion*

The performance evaluation demonstrates that the ONOS Command Line Interface imposes a smaller overhead to the Intent northbound interface in comparison to the RESTful interface and this behavior is consistent across all benchmarked Intent types. Our preliminary analysis indicates that the difference in performance is caused by both the extra computational cost of the RESTful interface and the cost associated with the remote connection used in our testbed environment for the RESful test application.

In addition to the above-mentioned performance results, we have also computed the maximum number of Intents a single ONOS instance was able to execute in our test topology. We ran 10 iterations of micro-benchmarks that submitted Intents in an infinite loop. For each iteration, we waited until ONOS was unable to install new Intents, and then computed for each execution the time required to install of all the intents of the same type. The results are summarized in Table VI.

TABLE VI
LAPTOP SPECIFICATIONS

| Intent type | Maximum number | Required time |
|---|---|---|
| Point-to-Point | 486,363.9 | 311,046.2 |
| Single-to-Multi-Point | 364,249.7 | 299,648.5 |
| Multi-to-Single-Point | 335,267.6 | 334,450.5 |

Finally, our experimental results show that ONOS can deliver higher performance if existing Intents types are extended to create more complex intents. This conclusion holds true for an ONOS-aware application that interacts with the SDN controller through the RESTful API. Our results also show that independently of the Intent type and the northbound interface used, at least for these basic Intents, the computational cost increases linearly with the number of Intents.

V. CONCLUSION AND FUTURE WORK

Our evaluation showed that while the RESTful interface is more flexible since it allows applications and services to directly request ONOS to update the network configuration, it has a greater overhead in comparison to the ONOS CLI, and consequently with any application directly linked with the ONOS core. As we have seen in our experiments, the RESTful interface is consistently more computationally costly than the CLI. When averaging this difference across all our experiments, we can conclude that the RESTful Intent code path is $9,488x$ slower than using the CLI. More specifically the average overall cost of using the RESTful interface for the *Point-to-Point* was $9,138$ times higher, $8,808$ times higher for the *Single-to-Multi-Point*, and $10,520$ times higher for the *Multi-to-Single-Point* Intent type than if using CLI.

For future work, we plan to create new complex Intents by using the basic Intents as building blocks and to evaluate their scalability and performance. These compound Intents will be formed by an external application requesting the network reconfiguration.

ACKNOWLEDGMENT

This work was supported in part by the Academy of Finland Project CSN under Grant No. 311654. The work was also supported in part by a direct funding from Nokia Bell Labs, Espoo, Finland.